\pdfoutput=1

\documentclass[11pt]{article}

\usepackage[]{ACL2023}

\usepackage{times}
\usepackage{latexsym}

\usepackage[T1]{fontenc}

\usepackage[utf8]{inputenc}

\usepackage{microtype}

\usepackage{inconsolata}
\usepackage{graphicx}

\usepackage{tabularx}
\usepackage{amssymb}
\usepackage[font=small,labelfont=bf]{caption}

\title{Local Life: Stay Informed Around You, A Scalable Geoparsing and Geotagging Approach to Serve Local News Worldwide}

\author{Deven Santosh Shah \\
  Microsoft  \\
  \texttt{devenshah@microsoft.com} \\\And
  Gosuddin Kamaruddin Siddiqi \\
  Microsoft  \\
  \texttt{gsiddiqi@microsoft.com} \\\And
  Shiying He \\
  Microsoft  \\
  \texttt{sylviahe@microsoft.com} \\\AND
  Radhika Bansal \\
  Microsoft  \\
  \texttt{rabansal@microsoft.com} \\}

\begin{document}
\maketitle
\begin{abstract}
Local news has become increasingly important in the news industry due to its various benefits. It offers local audiences information that helps them participate in their communities and interests. It also serves as a reliable source of factual reporting that can prevent misinformation. Moreover, it can influence national audiences as some local stories may have wider implications for politics, environment or crime. Hence, detecting the exact geolocation and impact scope of local news is crucial for news recommendation systems. There are two fundamental things required in this process, (1) classify whether an article belongs to local news, and (2) identify the geolocation of the article and its scope of influence to recommend it to appropriate users. In this paper, we focus on the second step and propose (1) an efficient approach to determine the location and radius of local news articles, (2) a method to reconcile the user's location with the article's location, and (3) a metric to evaluate the quality of the local news feed. We demonstrate that our technique is scalable and effective in serving hyperlocal news to users worldwide.


\end{abstract}

\section{Introduction}
\label{sec:introduction}
Local news is a vital source of information for users who want to stay updated on their surroundings or learn about other places. We follow the definition of local news published by \cite{shah2023whats}, which states:

\begin{quote} 
\begin{small}
\textit{We define local news articles that impact a specific set of users at the city/county/state level.}
\end{small}
\end{quote} 

\citet{shah2023whats} suggested there are two fundamental things that are required to keep the users informed about their surroundings, \textit{(1) identifying whether an article is local news, and (2) detecting and recognizing it's geolocation and the impact radius,} so that we could serve right news articles to the right audience. In this paper, we focus on the second step assuming the news articles we get are local.   

Detecting the article’s right location and showcasing it to the right audience not just improves user engagement \citep{robindro2017unsupervised} but it helps keep the community bind together, preserving it’s culture. It keeps the users informed of their neighborhood, may that be related to crime, events, new restaurants opening up, real estate, schools, sports, etc. Showcasing right local news to the cold start users is beneficial to start engaging them. Since, geolocation is the only information we get for cold start users, personalizing the feed based on the user's geolocation helps them engage better with the articles and convert them into warm users. For warm users, who have more preferences and behaviors, we can increase their retention rate by serving them high-quality local news.

We also review the existing literature on local news detection and geolocation extraction, and identify the main challenges and gaps in this research area. The papers (discussed in Section \ref{sec:relatedWork}) focus on extracting the geolocation information from the articles however, they fail to mention a way to reconcile the user's location with the article's location to serve the right local articles. The main challenges that we found in this research area are: 
\begin{enumerate}
\vspace{-9.2pt}

\item \textbf{Acronyms/Teams/Organizations/Highways as locations:} We came across multiple local news articles in which geographical location name wasn’t mentioned explicitly but was mentioned in the form of an acronym which might be a local place or a local organization, or a highway like I-5 or even a local sports team name. The locations of these acronyms are difficult to detect. For instance: "UW students voice concerns about recent University District crime"\footnote{https://www.msn.com/en-us/news/crime/uw-students-voice-concerns-about-recent-university-district-crime/ar-AA17wteu}, UW for University of Washington, "Seahawks have 2 of PFF's top 30 graded safeties for 2022"\footnote{https://www.msn.com/en-us/sports/nfl/seahawks-have-2-of-pffs-top-30-graded-safeties-for-2022/ss-AA17cqCH}, "Seahawks representing Seattle or infact entire Washington state. 

\item \textbf{Local news of broader interest:} We came across certain local news articles mentioning a specific city but could also be relevant to neighboring cities or regions. For instance: “Washington State Fair announces 2X Platinum artist with tickets on sale Wednesday” \footnote{https://www.msn.com/en-us/entertainment/news/washington-state-fair-announces-2x-platinum-artist-with-tickets-on-sale-wednesday/ar-AA16V9bk}. Even though Washington State Fair, Tacoma is mentioned in the article, people living in a radius of 50-80 miles of Tacoma would still love to see this article. Another example where local publishers capturing semi-local news articles like, "New WSU study says grocery stores can trick customers into spending more"\footnote{https://www.msn.com/en-us/money/companies/new-wsu-study-says-grocery-stores-can-trick-customers-into-spending-more/ar-AA17g8YC}

\item \textbf{Reconcile user’s location with the article’s location:} These techniques \citep{bell2015system, teitler2008newsstand, middleton2018location} do not mention how would they extract the geographical location of the user. There are certain techniques present that backtraces the IP address of the user to a location \citep{alt2010location}, however, reconciling these locations with the article’s location to serve the right local news would be a separate task. 

\item \textbf{Multiple locations present in the article:} We also came across multiple articles where there were multiple locations present. These approaches fail to handle this scenario. For instance: "No. 4 Arizona strives for best execution against offensively challenged Cal" \footnote{https://www.msn.com/en-us/sports/ncaabk/no-4-arizona-strives-for-best-execution-against-offensively-challenged-cal/ar-AA17fxR7}, this article mentions two states Arizona and California. People living in these two states would majorly be interested in seeing this article. 
\end{enumerate}

We showcase that not only our technique helps us understand the local publishers in an area but also scales worldwide and is effective in serving relevant hyperlocal news to the users globally. Our technique also helps in converting cold start users to warm users, and helps retain warm users.
 
Our major contributions include: \textit{(1) a novel efficient ensemble approach to detect and recognize the geographical location of the article and it’s impact radius, with high precision and high recall, (2) a technique to reconcile the user’s location with the geographical location of the article to serve the right local news, and (3) an evaluation metric to measure the quality of the local news feed.} We showcase that not only our technique helps us understand the local publishers in an area but also scales worldwide and is effective in serving relevant hyperlocal news to the users globally. Our technique also helps in converting cold start users to warm users, and helps retain warm users.

\section{Related Work}
\label{sec:relatedWork}

In the news domain, we realized that majority of the existing literature on geolocation extraction lie in one or more of the four issues buckets mentioned in section \ref{sec:introduction}. 

\citet{sankaranarayanan2009twitterstand} focuses on using user-generated content (UGC) platforms, such as Twitter, to gather breaking news in the area. They proposed a system that clusters tweets based on their geolocation name mentioned and the user's geolocation and assigns the geolocation foci of each cluster to all the tweets falling in it. However, this approach has several limitations: (1) it relies on user’s geolocation to extract the geolocation of the tweet, which may not always match the tweet’s geolocation, (2)  it ignores news impact radius, which depends on topic and audience, and (3) it requires manual identification of the users who post the news as tweets, which is hard to scale worldwide

\citet{tahmasebzadeh2021geowine} proposed a technique to use geolocation extracted from an image to showcase local news pertaining to that area. They proposed a technique that uses ResNets to predict the spatial geolocation and the entity type of an image, and then matches the entity name with the Wikidata corpus and the Event Registry to retrieve the relevant news articles. They restrict the entity types to 12 categories, such as landmarks, monuments, etc. However, this approach also has several limitations: (1) it misses other important local news categories, such as crime, sports, real estate, science, etc., (2) it depends on the quality and completeness of the Wikidata and the Event Registry, which may not be updated and accurate, and (3) it ignores user’s location and preferences, which may differ from the image’s location and entity type.

\citet{bell2015system} uses automatic speech recognition (ASR) to broadcast summaries of the news provided by the broadcasting channels. They proposed an ASR technique to generate the speech transcript and NER to extract and disambiguate geolocation with OpenStreet Map. However, they do not mention the serving technique for the generated local news summaries.

\citet{teitler2008newsstand} and \citet{middleton2018location} use a three-step process to get text geolocation: (1) geocoding, which maps the location names to the latitude and longitude coordinates, (2) geoparsing, which identifies the location names in text, and (3) geotagging, which assigns a geolocation to the text. They claim this approach can handle different text types and sources, and can be easily applied to the structured content like news articles present with news aggregators. However, this approach also has several limitations: (1) it depends on the size and scope of the gazetteer, which may not cover all location names or variations, (2) it ignores the impact radius of the local news articles, which may vary depending on the topic and the audience, and (3) they do not mention their approach to reconcile the article’s and user’s location to serve the right local news, which may differ in distance, relevance, and preference.

\citet{adar2017persalog}, inspired by personalizing the feed of users, proposed a tool that personalizes the content of the article based on two properties of the user: geolocation and demographics inference. They capture the location of the user by backtracing its IP address, and the location of the article by the journalist from a list of cities. They claim that this technique can provide more relevant and engaging news content to the user. However, this approach also has several limitations: (1) it restricts the location of the article to a list of cities, which may not cover all the possible local news sources or topics, and (2) it serves the local news content to the user only if they are within 50 miles of a nearby big city, which makes the technique difficult to scale worldwide.

\citet{gonccalves2021local} and \citet{vaataja2012location} focuses on using crowd-sourcing techniques to generate local news content from the citizens of an area. They proposed a framework to encourage local journalism in Portugal and Finland, respectively, by allowing the citizens to share photos, videos, posts about news-worthy events in their region, and local journalists to pick up the story if it seemed important. They claim that this technique can enhance the quality and diversity of local news content, and foster the community interaction and engagement. However, this approach also has several limitations: (1) it does not specify how to extract the location from the citizen’s posts, which may not always contain explicit or accurate location names, (2) it uses the user’s IP backtraced location as a proxy for the location of the post, which may not be the correct representative of the location of the event, and (3) it does not reconcile the user’s location and the post’s location to serve the right local news to the users, which may have different preferences and interests.

In contrast to these existing approaches, our proposed approach addresses all the issues and limitations mentioned above, and provides a more efficient and effective way to geoparse and geotag the location and radius of local news articles, and to match them with the user’s location and preferences. We also propose an evaluation metric to measure the quality and relevance of the local news feed, and demonstrate that our approach is scalable and effective in serving hyperlocal news to users worldwide.
\vspace{-5pt}
\section{Methodology}
We elaborated the local news serving problem into three subtasks, namely: (1) Geoparsing, detecting the geolocation and geotagging, recognizing it and determining the impact radius of the article, (2) reconciling the article and user's geolocation and preferences, and (3) delivering the personalized local news feed to the user.

\subsection{Geolocation as a form of geohash}
To match the user’s geolocation with the article’s geolocation, we use geohashes to represent both the user and the article. A geohash is a string of letters and numbers that encodes a rectangular area on the Earth’s surface. The longer the geohash, the smaller the area it covers. We use a single geohash to represent the user’s geolocation, and a set of geohashes to represent the article’s impacted geolocation. Table \ref{tab:geohash} shows the width and height of the rectangular areas corresponding to different geohash lengths\footnote{https://www.movable-type.co.uk/scripts/geohash.html}.

\begin{table}[htbp]
\centering
\resizebox{\columnwidth}{!}{%
\scriptsize\begin{tabular}{|l|l|l|}
\hline
\textbf{Geohash length} & \textbf{Cell width} & \textbf{Cell height} \\ \hline
1    $\leq$ & 5000 km    & 5000 km     \\
2 $\leq$ & 1250 km    & 625 km      \\
3 $\leq$            & 156 km     & 156 km      \\
4 $\leq$             & 39.1 km    & 19.5 km     \\
5 $\leq$             & 4.89 km    & 4.89 km     \\
6 $\leq$             & 1.22 km    & 0.61 km     \\
7 $\leq$             & 153 m       & 153 m        \\ \hline
\end{tabular}%
}
\caption{Rectangular geohash cell height and width coverage for geohashes of different lengths}
\label{tab:geohash}
\end{table}

\vspace{-20pt}
\subsection{Article Location detection}
Geoparsing and geotagging the geolocation and determining the impact radius of the article is a key task in local news serving. We use an ensemble of geoparsing and geotagging techniques for high precision and recall. The techniques we use are: (1) Location Table (LT) lookup, which detects and maps the location names to geohashes, (2) Bing Maps API, which detects and map the location names from the article and provides the bounding box coordinates to convert them into geohashes, and (3) Publisher-to-Location affinity mapping, which assigns geohashes to the article based on its publisher. For each article, we stamp geohashes of the article’s impacted geolocation by applying the following rules: (1) if the publisher of the article is in the publisher-to-location mapping, stamp mapped geohashes of length four, (2) if LT geohash and publisher-to-location geohash have same prefix of length two, stamp all LT geohashes of length four, (3) if BMA geohash and publisher-to-location geohash have same prefix of length two, stamp all BMA geohashes of length four, (4) if publisher-to-location mapping isn't available and if LT geohash and BMA geohash have same prefix of length two, stamp all LT and BMA geohashes of length four, (5) if publisher-to-location mapping and LT lookup are not available for the article, use BMA geohashes that are predicted with high confidence, and (6) if none of the rules apply, stamp no geohash to keep high precision over recall. We describe each of the techniques below:

\subsubsection{Location Table Lookup}
We compiled a list of locations in the United States, Canada, United Kingdom, and Australia. We checked if the article contained any location names or alias, and used the Bing Maps API bounding box coordinates to obtain geohashes of the location’s city-county/district-state-country geochain. We crowdsourced the evaluation of this technique using the UHRS hit apps \footnote{https://prod.uhrs.playmsn.com/UHRS/}, and got accuracy of 0.86 and recall of 0.38. However, we found that this technique is not scalable worldwide.

 \subsubsection{Bing Maps API Location}
There are several APIs that provide location information from a query/text, for instance, Google geocoding API\footnote{ https://developers.google.com/maps/documentation/geocoding}, OpenStreetMap Nominatim\footnote{http://wiki.openstreetmap.org/wiki/Nominatim} and Bing Maps API\footnote{https://www.microsoft.com/en-us/maps/choose-your-
bing-maps-api}. We use Bing Maps API. Bing Maps has an API to geoparse and geotag location information from text with confidence levels and location type. The confidence levels are High, Medium and Low, we use only those locations predicted with High and Medium confidence. Bing Maps API has different entity types, from Hospital, Building, Neighborhood to City, States, Monument and National Parks. We also use the coordinates information, like point coordinate of the location, which equates to the center point of the location and the bounding box coordinates that identifies four corners of a box like shape containing the location. We then extract all the geohashes within these coordinates. 

To judge the performance of the Bing Maps API on new articles, we generated recall metrics. As a ground truth, we assumed each news article is stamped with a location. To gain maximum recall we considered different parts of the articles, such as title, snippet, url and body and various combinations. (Table \ref{tab:rparts}). 

\begin{table}[htbp!]
\scriptsize\begin{tabular}{|p{0.7cm} p{0.35cm} p{0.6cm} p{0.3cm} p{0.3cm} p{0.5cm} p{0.8cm} p{0.8cm}|}
\hline
\textbf{Market} & \textbf{Title} & \textbf{Snippet} & \textbf{URL} & \textbf{Body} & \textbf{Title + Body} & \textbf{Title + Snippet + URL} & \textbf{Title + Snippet + Body} \\ \hline
\textbf{en-au}  & 0.75          & 0.45            & 0.75        & 0.73       & 0.88               & 0.85                         & \textbf{0.90}                  \\ 
\textbf{en-ca}  & 0.73       & 0.54         & 0.72      & 0.73      & 0.88              & 0.83                      & \textbf{0.89}               \\ 
\textbf{en-gb}  & 0.59       & 0.31         & 0.58     & 0.55      & 0.69              & 0.69                      & \textbf{0.74}               \\ 
\textbf{en-us}  & 0.72       & 0.66         & 0.69      & 0.75      & 0.91               & 0.86                      & \textbf{0.94}               \\ \hline
\end{tabular}
\scriptsize\caption{ Comparing BMA Location Recall Metrics for news article attributes; Snippet extracted by in-house trained textrank model; Title, snippet, body trimmed to starting and ending 10 words each for QPS reduction.}
\label{tab:rparts}
\end{table}
        
We found that the Title + Snippet + Body query gave the best precision and recall for the BMA location. We also noticed that URL in query could bias the BMA to the location name in the provider’s name, which might not be article's location. For example, KOMO-TV Seattle reported a news from Sammamish area, for instance: "Person shot during home invasion in Sammamish"\footnote{https://www.msn.com/en-us/news/crime/person-shot-during-home-invasion-in-sammamish/ar-AA17qJG0}, but the BMA might pick Seattle as the article’s location, which would prevent the article from reaching the Sammamish users, and also annoy the Seattle users who might not care about Sammamish news. Therefore, we decided to exclude the URL from the query for BMA, thus avoiding label bias \citep{shah2019predictive}. 

We measured the precision of the recalled locations with crowd-sourced evaluation using the UHRS hit apps. For each article, we shared, the title, snippet, body and made the URL available for the user to read the article. We then asked the users if detected location is correct for the given article. These numbers aggregated by different locales are listed in Table \ref{tab:prbma}.
\begin{table}[htbp]
\centering
\resizebox{\columnwidth}{!}{
\begin{tabular}{|l|l|l|l|}
\hline
\textbf{Market} & \textbf{Baseline Recall} & \textbf{BMA Recall} & \textbf{BMA Precision} \\ \hline
en-au            & 0.35     & 0.80 & 0.94      \\
en-gb & 0.18    & 0.84 & 0.97      \\
en-ca            & 0.25     & 0.85 & 0.93      \\
en-us &  0.66   & 0.84 & 0.91      \\
de-de             & 0.41  & 0.73 & 0.95        \\
es-es             & NA    & 0.69 & 0.96     \\
fr-fr             & NA    & 0.63 & 0.94     \\
it-it             & NA    & 0.59 & 0.85     \\
ja-jp             & NA    & 0.83 & 0.84    \\ 
zh-cn             & NA    & 0.86 & 0.92     \\\hline
\end{tabular}
}
\caption{Precision-Recall Metrics comparison between baseline and BMA's extraction model.}
\label{tab:prbma}
\end{table}

\vspace{-22pt}
\subsubsection{Publisher-Location affinity}
\begin{figure}[tbh!]
    \centering
    \includegraphics[width=\columnwidth]{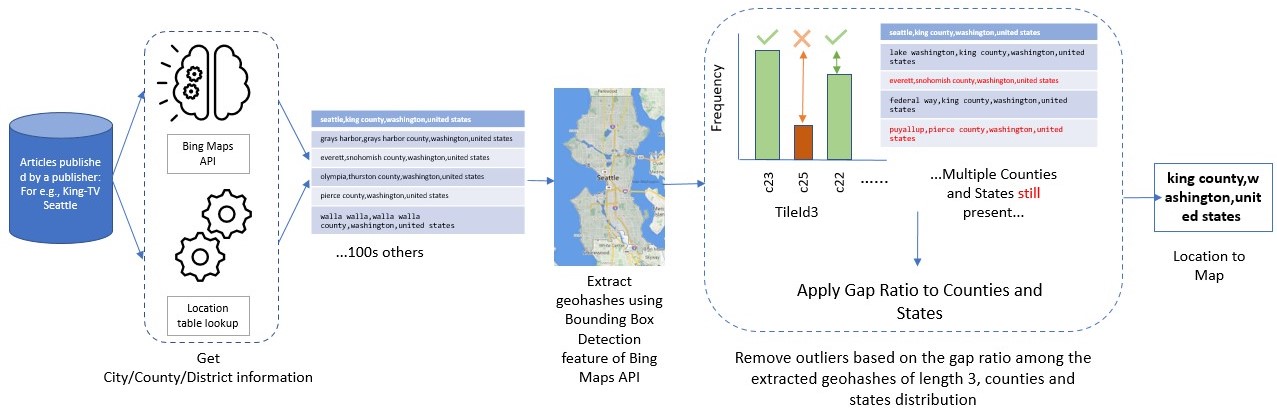}
    \caption{\textit{Publisher-to-Location Affinity}: Illustrating the steps to obtain the local publisher-to-location affinity.}
    \label{fig:publocaff}
\end{figure}

 We observed that some local news articles did not explicitly mention the location name in them, such as "Boba exhibit opening at the Chinese American Museum"\footnote{https://www.msn.com/en-us/video/peopleandplaces/boba-exhibit-opening-at-the-chinese-american-museum/vi-AA17g8k4}, or "Eastbound 520 Bridge Closure Planned Saturday Night"\footnote{https://www.msn.com/en-us/news/us/eastbound-520-bridge-closure-planned-saturday-night/ar-AA17gizN}, or "Warriors win against OKC reveals silver lining to Steph Curry’s absence"\footnote{https://www.msn.com/en-us/sports/nba/warriors-win-against-okc-reveals-silver-lining-to-steph-curry-s-absence/ar-AA17dbSv}which were published by local news providers like “CBS Los Angeles”, “Patch”, and “Mercury News” respectively. We also noticed that users in nearby cities, such as Seattle and Bellevue, will be interested in each other’s local news, and local news providers often covered news from neighboring cities as well. To handle these cases, we identified the strongly local providers using the method proposed by \citet{shah2023whats}. We then computed the publisher-to-location affinity for these providers using all the LT and BMA locations recognized on their articles in a month. We used the gap ratio \citep{shah2023whats} to filter out the remote locations that the local providers rarely covered, and retained the locations that they frequently covered. We assumed that the articles that did not have the location name in them were from these frequent locations, which could be at the city/county/district level. For example, we mapped the publisher “KOMO-TV Seattle” to the location “King County, Washington, United States”, suggesting that KOMO-TV Seattle mainly covers news articles from King County. We created a mapping of publisher-to-location affinity using the following steps, which are also illustrated in Figure \ref{fig:publocaff}:

\begin{itemize} 
    \item Collect all the articles from a specific provider in a given time range.
    \item Apply Bing Maps API and Location Table lookup to extract the locations from these articles.
    \item Use Bing Maps API to get the geohashes from the bounding box of the extracted locations.
    \item Use gap ratio on geohashes of length three to filter out the outlier geohashes, and map the remaining geohashes of length four to their corresponding locations \item Use gap ratio to filter out outlier counties and states.
    \item The remaining locations are the high-affinity locations that the provider covers.
\end{itemize}

The mapped location is converted to a set of geohashes using the Bing Maps API bounding box coordinates. The publisher-to-location affinity mapping helped address the issues of \textbf{"Acronyms as locations"}, and \textbf{"Local news of broader interest"} issues discussed in the Introduction section.

\subsection{Local news serving}
We obtain the user location from various sources, such as their IP address or their input on the weather card. We convert this location, especially the latitude and longitude, to a geohash of length four, which is used to retrieve the articles to show to the users. If there are not enough articles in that geohash, we backfill it with the nearest popular city, assuming that some users might commute to big cities for work and want to stay informed about them. The retrieved articles are then ranked by an in-house trained ranker to display to the users. Converting geolocations to geohashes helped reconcile the user's and article's location to correct serving.

\section{Evaluation Metrics}
We evaluated our End-To-End Local News serving technique, by performing an online A/B experiment. We measured our technique specifically with the earth's distance between the user's location and identified document location. In addition to this, we accounted the fact that if a user's location for example a city Seattle, Washington matches with document's location, for example the same city, we assume the distance as 0 kms. This suggests our ability to serve hyper-local news at the most granular level such as a city. Similarly, we take into account for other geographical divisions such as county and state.

\section{Results}
\begin{figure}[htbp]
    \centering
    \includegraphics[width=\columnwidth]{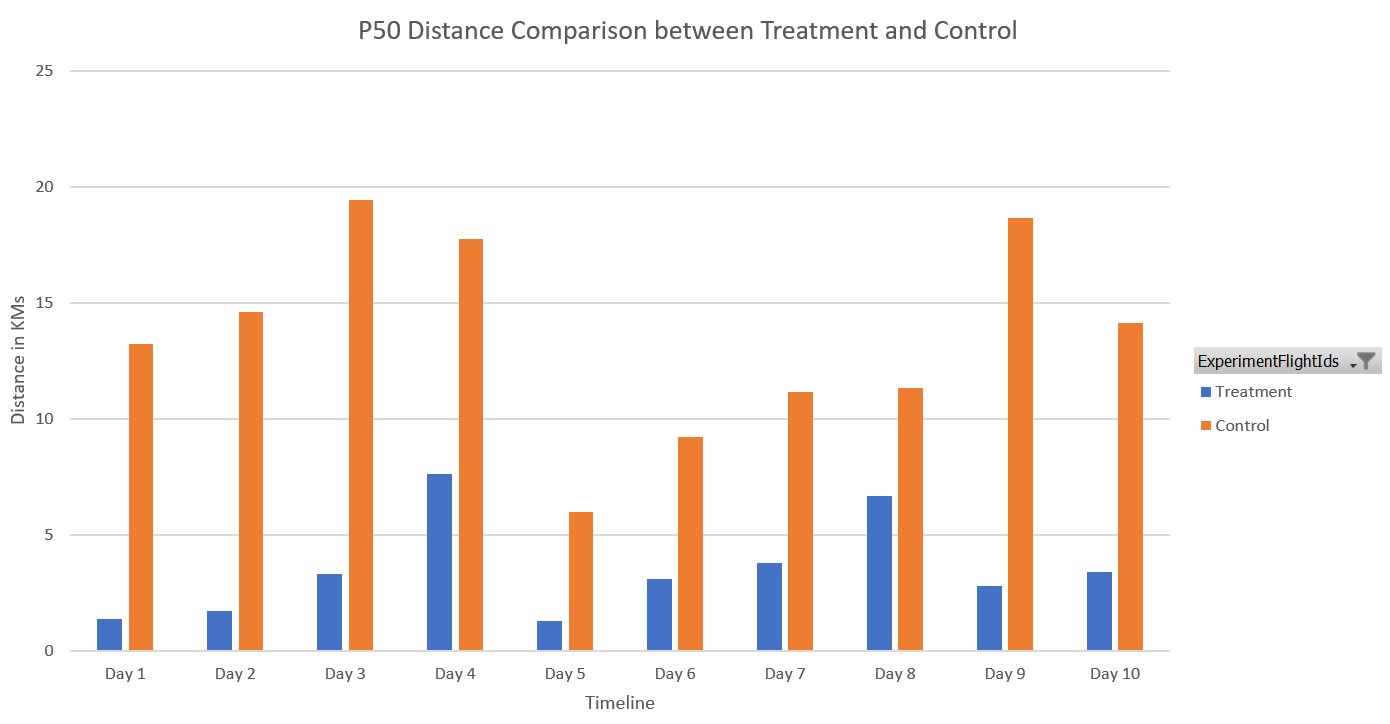}
    \caption{\textit{P50 Distance Metrics}: Comparing P50 metrics between treatment with Local News serving technique and Control }
    \label{fig:p50Metrics}
\end{figure}

\begin{figure}[htbp]
    \centering
    \includegraphics[width=\columnwidth]{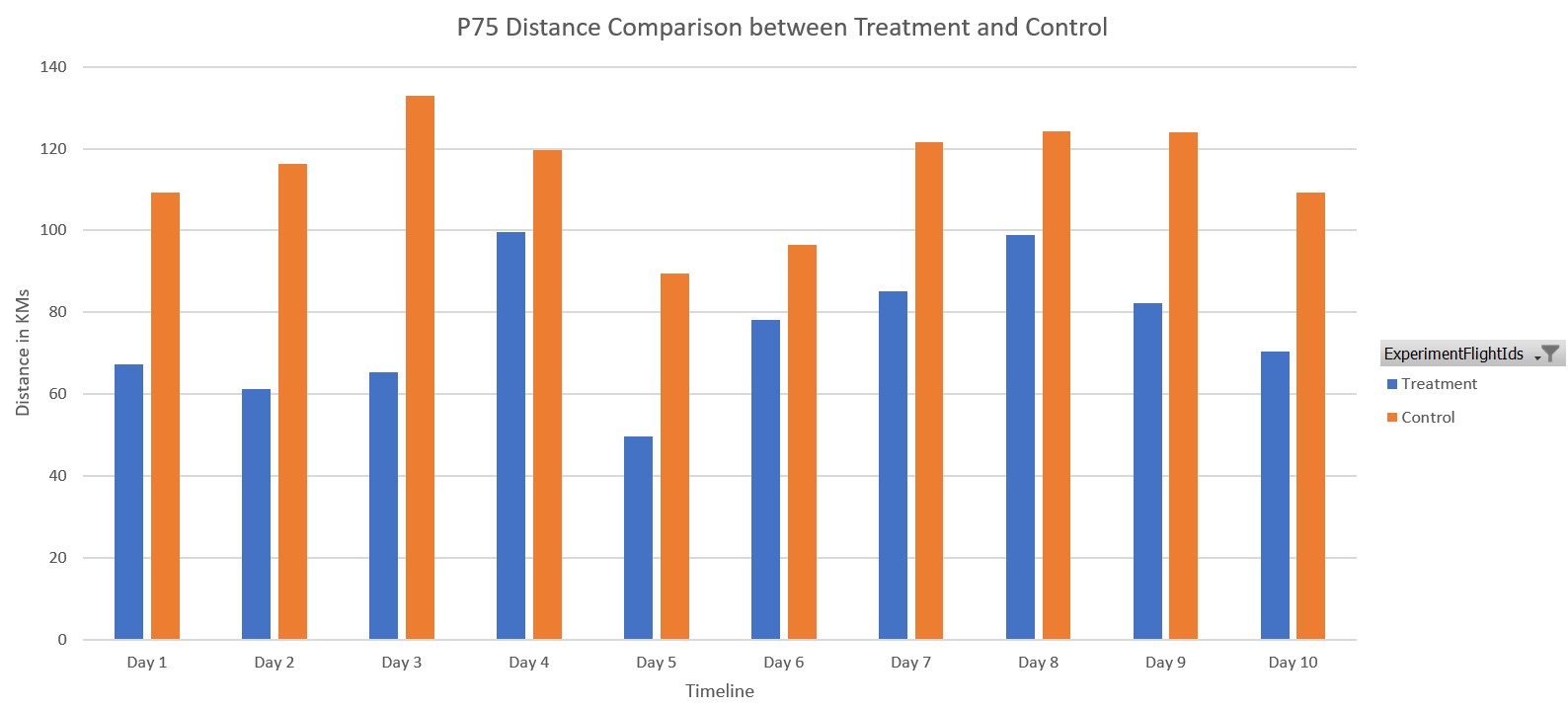}
    \caption{\textit{P75 Distance Metrics}: Comparing P75 metrics between treatment with Local News serving technique and Control }
    \label{fig:p75Metrics}
\end{figure}

For our baseline, we used the publisher-to-DMA mapping. We manually mapped strongly local publishers to a DMA, and stamped all the articles from the local publisher with all the geohashes corresponding to the mapped DMA. DMA stands for Designated Market Area, which is a geographical region comprising a group of counties and zip codes defined by Neilsen \footnote{https://markets.nielsen.com/us/en/contact-us/intl-campaigns/dma-maps/}. There are 210 DMAs in the United States, mostly used by the television and radio channels to broadcast to a set of users. We used lexicon-based knowledge constraints \citep{shah2021distantly} to tag DMA-based geohashes on the article. These lexicons were manually created. We compared our ensemble technique of detecting the article’s location to the DMA tagging, to measure the localness of the delivered local news articles in terms of distance, and to ensure the relevance of the local news shown to the users.
We conducted on online experiment to compare the distance between users and document. Treatment group was exposed to End to End Local News serving technique and we observed improvement on 50\^th and 75\^th Percentile distance. 50\^th Percentile distance improved from 15 kms on average to 8 kms on average (Fig:\ref{fig:p50Metrics}) and 75\^th Percentile distance improved from 120 kms on average to 80 kms on average (Fig:\ref{fig:p75Metrics}). We also observed an improvement in content interactions per user session for cold as well as warm start users.

\section{Conclusion}
In this paper, we proposed (1) an ensemble approach to determine the geographical location of the article and it's impact radius, (2) a technique to reconcile the user's location with the geographical impacted location of the article to serve the right local news, and (3) an offline and online evaluation metrics to measure the quality of the local news feed. We also showcased our technique resolves the major issues with local news to showcase right local articles to the right set of audience. We eventually showcased that our technique is scalable worldwide, and helps convert cold start users to warm start users.

\section*{Limitations}
Currently we have a rule based ensemble approach to stamp geolocation on an article. There is a good potential to train a ML model to make the selection from the models that provide the geolocation. Bing Maps API is a query based API, and hence one limitation that we currently have in our system is we use starting and ending 10 words to save on the QPS. Having an in-house or off-the-shelf NER model to detect the location mention in the article and pass it as an input to the Bing Maps API would help increase the precision of the BMA Location extraction. 

\bibliography{local}
\bibliographystyle{acl_natbib}

\end{document}